\documentclass[aps,prb,twocolumn,superscriptaddress,showpacs]{revtex4-1}
\usepackage{graphicx}
\usepackage{dcolumn}
\usepackage{bm}
\usepackage{amsfonts}
\usepackage{amsmath}
\usepackage{color}

\definecolor{rodcolor}{rgb}{0.0, 0.8, 0.0}
\definecolor{rodcolor2}{rgb}{0.0, 0.4, 0.4}

\begin{document}
\title{Coherent edge mixing and interferometry in quantum Hall bilayers}
\author{Stefano Roddaro}
\affiliation{NEST, Istituto Nanoscienze-CNR and Scuola Normale Superiore, I-56126 Pisa, Italy}
\affiliation{Istituto Officina dei Materiali CNR, Laboratorio TASC, Basovizza, I-34149 Trieste, Italy}
\author{Luca Chirolli}
\affiliation{NEST, Scuola Normale Superiore and Istituto Nanoscienze-CNR, I-56126 Pisa, Italy}
\author{Fabio Taddei}
\affiliation{NEST, Istituto Nanoscienze-CNR and Scuola Normale Superiore, I-56126 Pisa, Italy}
\author{Marco Polini}
\affiliation{NEST, Istituto Nanoscienze-CNR and Scuola Normale Superiore, I-56126 Pisa, Italy}
\author{Vittorio  Giovannetti}
\email{v.giovannetti@sns.it} 
\affiliation{NEST, Scuola Normale Superiore and Istituto Nanoscienze-CNR, I-56126 Pisa, Italy}
\begin{abstract}
We discuss the implementation of a beam splitter for electron waves in a quantum Hall bilayer. 
Our architecture exploits inter-layer tunneling to mix edge states belonging to different layers. 
We discuss the basic working principle of the proposed coherent edge mixer, possible interferometric implementations based on existing semiconductor-heterojunction technologies, 
and advantages with respect to canonical quantum Hall interferometers based on quantum point contacts. 
\end{abstract}
\maketitle
\section{Introduction}
\label{sect:intro}

Chiral edge states living at the boundary of a two-dimensional (2D) quantum Hall (QH) phase~\cite{generalbookQHE,Giuliani_and_Vignale} constitute a fascinating playground both for the investigation of 
fundamental properties of one-dimensional electron liquids~\cite{chang_rmp_2003} and the implementation of innovative devices. 

A number of novel QH electron interferometers have been demonstrated in recent years, opening a new window of investigation on coherent electron transport in solid-state devices. In particular, edge ``beams'' have been exploited for the realization of a variety of electronic interferometers reproducing the Mach - Zehnder~\cite{Ji2003,Roulleau2007,Litvin2008}, Fabry - Perot~\cite{McClurePRL2009} and Hanbury Brown - Twiss~\cite{HennyScience1999,Neder2007} schemes adopted in optics. These can have an important impact both on the fundamental investigation of quantum transport phenomena of electrons in solids and as possible implementations for quantum computing~\cite{qcompt}. In these circuits, mixing between edge states has so far been achieved using beam splitters (BSs) based on quantum point contacts (QPCs). Fascinating but still puzzling phenomena have been highlighted in these devices~\cite{Schneider2011}, in particular in relation to finite-bias visibility and edge reconstruction phenomena. While the QPC approach has proven successful, the intrinsic geometry of this BS implementation makes it necessary to adopt non-simply connected 2D electron gases (EGs) and limits the complexity and size of the achievable 2D 
QH circuits. In addition, such BS structures represent a potentially complex circuit element displaying non-linear characteristics~\cite{Roddaro2005,Roddaro2009} as well as, in some cases, fractional substructures~\cite{Paradiso2012} which can have an impact on the overall behavior of the 2D QH circuit. Alternative interesting device schemes exist and are based on tunneling between co-propagating edge modes~\cite{GiovannettiPRB2008, Karmakar2011,Paradiso2011,Chirolli2012,Deviatov2011}.

\begin{figure}[h]
\begin{center}
\includegraphics[width=0.4\textwidth]{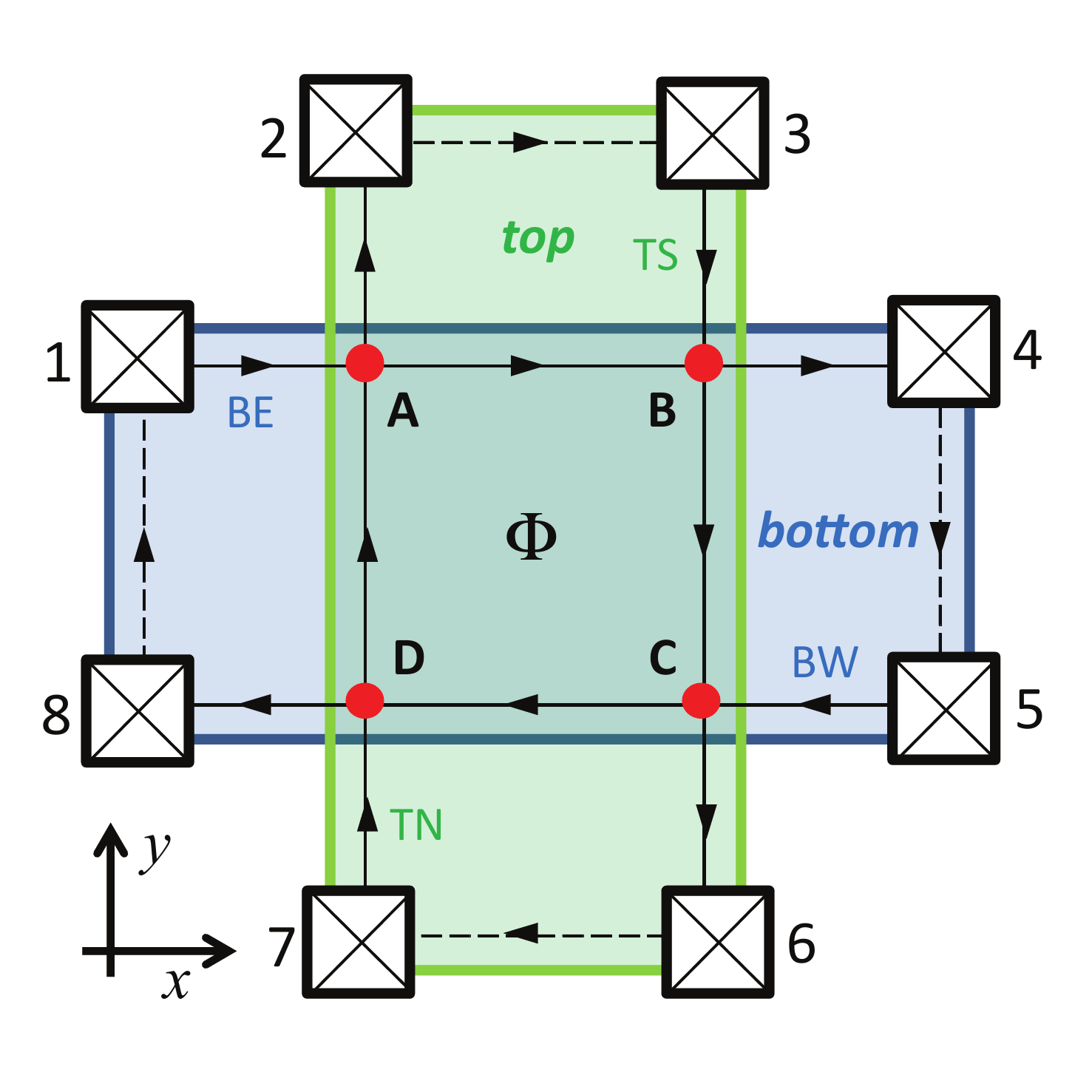}
\caption{(Color online) A top view of a quantum Hall bilayer interferometer (QHBI). In this cartoon, the QHBI is obtained by employing two overlapping orthogonal two-dimensional electron gases (``top" and ``bottom" conducting strips of width $W$) in the integer quantum Hall regime. The two subsystems are vertically separated by a distance $h$. We assume that only inter-layer tunneling couples the two subsystems. In this configuration, transport takes place only via the quantum Hall edge states of the two subsystems. Inter-layer tunneling is effectively active only at the edge crossing points, labeled by A, B, C, and D.\label{fig:one}}
\end{center}
\end{figure}

Here, we discuss a different paradigm to edge-beam interferometry, which is based on the exploitation of a {\it QH bilayer}~\cite{generalbookQHE}. This is a system composed of two closely-spaced 2DEGs. We assume that each of the two 2DEGs is in the QH regime and has edge states. Our BS for electron waves is based on inter-layer tunneling (Fig.~\ref{fig:one}), {\it i.e.} hybridization between the two 2DEGs. 
We show that momentum-conserving tunneling between the two 2DEGs can be used to scatter an edge mode localized in one layer into an edge mode localized into the other in a controllable way. In particular, when edge modes in the two 2DEGs cross at a finite angle, the scattering process effectively involves only a region of the order of the magnetic length, $\ell_B=\sqrt{\hbar c/(eB)}$, around the geometrical crossing point of the edge modes. Using this approach, many interferometric schemes can be achieved by employing relatively simple gating geometries.

This Article work is organized as follows. In Section~\ref{sect:interferometer} 
we describe a basic interferometric scheme, which can be obtained by overlapping two orthogonal two-dimensional electron gases in a bilayer system. Section~\ref{sect:eigenvalueproblem} describes the formal scattering problem ruling the interferometer behavior. Section~\ref{sect:exactsolution} discusses the adopted scattering matrix for the beam splitter and analyzes the details of its microscopic origin. Finally, in Sect.~\ref{sect:conclusions} we summarize our main findings and draw our conclusions.

\section{Interferometer layout}
\label{sect:interferometer}

We consider a quantum Hall bilayer interferometer (QHBI) constituted by two two-dimensional electron gases (2DEGs) (the bottom layer, B, and the top layer, T) extending in the $x,y$ plane and separated by a distance $h$ along the vertical direction $z$. Both 2DEGs are assumed to be in the integer quantum Hall regime, induced by the presence of a (uniform) quantizing magnetic field ${\bm B}$ pointing along the $z$-axis, {\it i.e.} ${\bm B} = B {\hat {\bm z}}$. We assume that the two subsystems are coupled only by a uniform tunneling term of strength $\Delta_{\rm SAS}$. While $\Delta_{\rm SAS}$ is mostly determined by the heterostructure design (height of the barrier, inter-layer separation $h$, etc) it can also be tuned to some extent by an in-plane magnetic field~\cite{Hu92}. In principle, the two 2DEGs are also coupled by electron-electron interactions. In this Article, however, we neglect these effects and discuss the working principle of our QHBI at the single-particle level. The inclusion of inter-layer electron-electron interactions is definitely challenging and expected to be responsible for an interesting phenomenology, which is, however, well beyond the scope of the present Article.

We further assume 
that the electrostatic potentials that are responsible for the lateral confinement within each layer, are ``layer-dependent'' and characterized by mutually orthogonal ``longitudinal'' directions. Specifically, introducing  the in-plane coordinate vector ${\bm r}=x\hat{\bm x}+y \hat{\bm y}$, we take the confining potential $V_{\rm B}({\bm r})$ in the B layer to be translationally invariant with respect to $x$, while the one 
$V_{\rm T}({\bm r})$ in the  T layer to be translationally invariant with respect to $y$, {\it i.e.}
\begin{eqnarray} \label{confpot}
V_{\rm B}({\bm r}) = V_{\rm B}(y) \;, \qquad V_{\rm T}({\bm r}) = V_{\rm T}(x)~.
\end{eqnarray}
The specific functional form of $V_{\rm B}(y)$ and $V_{\rm T}(x)$ will be fixed later. Under this condition and at low energies, charge transport is dominated by single-particle chiral-edge modes propagating 
along the $\hat{\bm x}$ axis in the B layer, and along the $\hat{\bm y}$ axis in the T layer, granting the setup the form of two rectangular Hall bars of width $W$, which are crossing perpendicularly as schematized in Fig.~\ref{fig:one}. Analogous configurations have been realized experimentally in semiconductor-heterojunction double quantum wells thanks to top and bottom gating and to inter-layer screening effects~\cite{Sandia,Klitzing, Eisenstein,Eisenstein2}. These allow depleting one of the two quantum wells without substantially altering the other and thus can lead to independent carrier-density profiles in T and B layers. 
Significant tunneling can be obtained for suitably designed barriers.

\subsection{Model Hamiltonian} 

The Hamiltonian of our system consists  in the sum of two terms,  $\hat{\cal H}=\hat{\cal H}_0+ \hat{\cal H}_{\rm tun}$, which describe respectively the free evolution of the electrons in the two layers and the  inter-layer tunneling  coupling. Introducing  Fermionic field operators $\hat{\Psi}_{\sigma}({\bm r})$ that satisfy canonical anti-commutations rules~\cite{NOTA}, the first contribution can be expressed as
\begin{eqnarray}\label{Eq:H0}
  \hat{\cal H}_0 &=& \sum_{\sigma= {\rm T}, {\rm B}} \int d^2{\bm r}~\hat{\Psi}^\dagger_{\sigma}({\bm r}) H_{\sigma}({\bm r})
  \hat{\Psi}_{\sigma}({\bm r})~,
  \end{eqnarray}
where
\begin{equation}\label{ham1particle}
H_{\sigma}({\bm r})=\frac{1}{2 m^*} \left[ - i \hbar \nabla_{\bm r} + \frac{e}{c}{\bm A}({\bm r})\right]^2 +V_{\sigma}({\bm r})~.
\end{equation}
Here  $m^*$ is the electron band mass ($m^* \sim 0.067~m_{\rm e}$, for example, for GaAs, where $m_{\rm e}$ is the electron mass in vacuum) and ${\bm A}({\bm r})$ is the vector potential associated with the uniform magnetic field ${\bf B}=B\hat{\bf z}$. We work in the symmetric gauge, ${\bm A}({\bm r})= ({\bf B} \times  {\bm r})/2$. The inter-layer tunneling term can be written as~\cite{generalbookQHE} 
 \begin{equation} \label{INT}
  \hat{\cal H}_{\rm tun}=- \frac{\Delta_{\rm SAS}}{2} 
  \int d^2{\bm r}~\left[{\hat \Psi}^\dagger_{\rm T}({\bm r}) {\hat \Psi}_{\rm B}({\bm r}) +{\rm H.}{\rm c.}\right]~,
  \end{equation}
where $\Delta_{\rm SAS}$ is the symmetric-to-antisymmetric tunneling gap. Values of $\Delta_{\rm SAS}$ ranging from $\ll 1~\mu{\rm eV}$~\cite{Spielman} up to the meV scale~\cite{Luin} have been demonstrated in literature by tuning the barrier design and the in-plane magnetic field. 
 
Equations~(\ref{Eq:H0}) and (\ref{INT}) can be casted in a more compact form by expressing them in terms of the eigenstates of the ``unperturbed" problem ({\it i.e.} $\Delta_{\rm SAS} = 0$). To this end, we introduce the eigenvalues $\epsilon_{\sigma,n,k}$ of the single-particle Hamiltonians $H_\sigma({\bm r})$ and the corresponding eigenfunctions $\phi_{\sigma,n,k,}({\bm r})$:
  \begin{eqnarray}
  \phi_{{\rm B} ,n,k}({\bm r})&=&e^{-ixy/(2\ell_B^2)}\frac{e^{ikx}}{\sqrt{2\pi}}\; \chi_{{\rm B},n,k}(y)~, \nonumber \\
  \phi_{{\rm T},n,k}({\bm r})&=&e^{ixy/(2\ell_B^2)}\frac{e^{iky}}{\sqrt{2\pi}}\;\chi_{{\rm T}, n,k}(x)~, \label{Eq:WFB}
  \end{eqnarray}
where $k\in \mathbb{R}$  is the eigenvalue of the magnetic translation operator along the ${\hat {\bm x}}$ (${\hat {\bm y}}$) axis for the B (T) layer, while $n\geq 0$  is a discrete index which labels the Landau levels. The functions $\chi_{\sigma,n,k}(u_\sigma)$ determine 
the transverse structure of the propagating modes (\ref{Eq:WFB}) and are the eigenfunctions of the transverse Hamiltonian:
  \begin{equation}
  H^{\rm tr}_{\sigma,k}(u_\sigma)=-\frac{\hbar^2}{2m^*}\frac{\partial^2}{\partial u^2_\sigma} +\frac{m^* \omega^2_{\rm c}}{2}(u_\sigma - {\bar u}_{\sigma,k})^2+V_\sigma(u_\sigma)~.
  \end{equation}
Here $\omega_{\rm c}= eB/(m^* c)$ is the cyclotron frequency, ${\bar u}_{{\rm B},k}= -{\bar u}_{{\rm T},k} = k \ell^2_B$, and $u_\sigma$ is the ``transverse" coordinate:
\begin{equation}
u_{\rm B} = y~, \qquad \qquad u_{\rm T} = x~.
\end{equation}

Since the functions  $\phi_{\sigma,n,k}({\bm r})$ form a complete orthonormal set~\cite{NOTA}, we can use them to expand the operators $\hat{\Psi}_\sigma({\bm r})$, obtaining the identities 
  \begin{eqnarray} \label{expansion}
  \hat{\Psi}_{\sigma}({\bm r})&=&\sum_{n=0}^{\infty} \int dk~\phi_{\sigma,n,k}({\bm r})\; \hat{c}_{\sigma,n,k}\;,\\
  \hat{c}_{\sigma,n,k}&=&  \int d^{2}{\bf{r}} \; \phi^*_{\sigma,n,k}({\bm r})\; \hat{\Psi}_{\sigma}({\bm r})\;,
  \end{eqnarray}
where  $\hat{c}_{\sigma,n,k}$  are the Fermionic annihilation operators associated with the chiral-edges modes of the unperturbed Hamiltonian $\hat{\cal H}_0$ that satisfy the eigenmode equation 
  \begin{eqnarray} \label{eigenmodeeq}
  [\hat{\cal H}_0 , \hat{c}_{\sigma,n,k} ] = - \epsilon_{\sigma,n,k} \; \hat{c}_{\sigma,n,k}~.
  \end{eqnarray}
Analogously, we can write
  \begin{eqnarray} \label{H0}
  \hat{\cal H}_0 =\sum_{\sigma={\rm B,T}}\sum_{n=0}^\infty \int dk~\epsilon_{\sigma,n,k}\; \hat{c}^{\dag}_{\sigma,n,k}\hat{c}_{\sigma,n,k}~,
  \end{eqnarray}
and 
\begin{eqnarray} \label{coupling1}
\hat{\cal H}_{\rm tun}=\sum_{n,n'}\int dk dk'\left[t_{n',k';n,k}\hat{c}^{\dag}_{{\rm T},n',k'}\hat{c}_{{\rm B},n,k}+{\rm H.c.}\right]~,
\end{eqnarray}
with the tunnel matrix elements defined as
\begin{eqnarray}
t_{n',k';n,k}&=&-\frac{\Delta_{\rm SAS}}{2}\int d^2{\bm r}\; \phi^*_{{\rm T},n',k'}({\bm r})\phi_{{\rm B},n,k}({\bm r})\nonumber \\
&=& -\frac{\Delta_{\rm SAS}}{4\pi}e^{-ikk'\ell^2_{\rm B}}{\cal F}_{n',n}(k',k)~, \label{tunn1}
\end{eqnarray}
where ${\cal F}_{n',n}$ is a form factor that describes the overlap between transverse wavefunctions residing in different layers: 
\begin{eqnarray}
{\cal F}_{n',n}(k',k)&=&\int dxdy~e^{-i(x+k' \ell_B^2)(y-k\ell_B^2)/\ell_B^2}\nonumber \\
&&\qquad \quad \times \;\; \chi_{{\rm T},n',k'}^*(x)\chi_{{\rm B},n,k}(y)\;. \label{FormFactor}
\end{eqnarray}  
\section{The eigenvalue problem in the presence of tunneling} 
\label{sect:eigenvalueproblem}

To characterize how inter-layer tunneling affects the transport properties of the system at hand we need to study  the eigenmode equation 
  \begin{eqnarray}\label{eigenmode}
  [ \hat{\cal H}, \hat{\gamma}(\epsilon)] = - \epsilon \; \hat{\gamma}(\epsilon)~,
  \end{eqnarray}  
which plays the role of Eq.~(\ref{eigenmodeeq}) when tunneling is taken into account. More explicitly, Eq.~(\ref{eigenmode}) can be written for the components of the pseudospinor 
wavefunction ${\psi}^{(\epsilon)}({\bm r})=( \psi_{{\rm B}}^{(\epsilon)}({\bm r})  , \psi_{{\rm T}}^{(\epsilon)}({\bm r}))^\top$, {\it i.e.} 
  \begin{eqnarray}
  {H}_{\rm B}({\bm r})  \psi_{{\rm B}}^{(\epsilon)}({\bm r})  
  -\frac{\Delta_{\rm SAS} }{2} \psi_{\rm T}^{(\epsilon)} ({\bm r}) &=&  \epsilon\; \psi_{\rm B}^{(\epsilon)} ({\bm r})\;,
  \nonumber \\
  {H}_{\rm T}({\bm r})  \psi_{{\rm T}}^{(\epsilon)}({\bm r})   \label{eqspin}
  -\frac{\Delta_{\rm SAS} }{2} \psi_{\rm B}^{(\epsilon)} ({\bm r}) &=&  \epsilon \; \psi_{\rm T}^{(\epsilon)} ({\bm r})\;,
  \end{eqnarray} 
which make it explicit that  the eigenfunctions of the full Hamiltonian are delocalized in the two layers due to the tunneling coupling $\Delta_{\rm SAS}$. 
Both~(\ref{eigenmode}) and (\ref{eqspin}) can also be casted as 
  \begin{eqnarray} \nonumber 
  (\epsilon - \epsilon_{{\rm B}, n,k}) \alpha_{{\rm B}, n,k}^{(\epsilon)} &=& \sum_{n'} \int dk'  \; t^*_{n',k';n,k} \;
  \alpha_{{\rm T},n',k'}^{(\epsilon)}\;,  \\
  ( \epsilon - \epsilon_{{\rm T}, n, k})  \alpha_{{\rm T}, n,k}^{(\epsilon)} &=& \sum_{n'} \int dk'  \;
  t_{n',k';n,k} \;\alpha_{{\rm B}, n',k'}^{(\epsilon)}   \;, \nonumber \\ \label{eigenEqint}
  \end{eqnarray} 
where  $\alpha_{\sigma, n,k}(\epsilon)$ are complex coefficients relating $\hat{\gamma}(\epsilon)$ to its unperturbed counterpart $\hat{c}_{\sigma,n,k}$ and the components of the spinor ${\psi}^{(\epsilon)}({\bm r})$ to the functions $\phi_{{\sigma},n,k,}({\bm r})$ defined in Eq.~(\ref{Eq:WFB}), {\it i.e.}
  \begin{eqnarray}
  \hat{\gamma}(\epsilon) &=& \sum_{\sigma = {\rm T}, {\rm B}} \sum_n \int dk \; [\alpha_{\sigma, n,k}^{(\epsilon)}]^* \; \hat{c}_{\sigma,n,k}   
  \;,\label{unitarymap}\\
  {\psi}^{(\epsilon)}_\sigma({\bm r})&=& \sum_n \int dk \; \alpha_{\sigma, n,k}^{(\epsilon)} \;  \phi_{{\sigma},n,k,}({\bm r})\;. \label{eigenfuntionint}
  \end{eqnarray} 
At this point it is clear that the total field operator $\hat\Psi({\bm r})=\sum_\sigma\hat\Psi_\sigma({\bm r})$ can be expanded in the basis of the eigenvectors $\hat\gamma(\epsilon)$ as
  \begin{equation}
  \hat\Psi({\bm r})=\sum_\epsilon \; \psi^{(\epsilon)}({\bm r})\; \hat\gamma(\epsilon)~,
  \end{equation}
where $\epsilon$ compactly labels all quantum numbers of the problem (\ref{eqspin}). 

The eigenmode equation~(\ref{eqspin})  defines a scattering process in which an edge beam described by the 
unperturbed energy eigenstate~(\ref{Eq:WFB}) and propagating in the B layer, say, is partially transmitted into the T layer. The associated probability amplitudes strongly depend on the specific choice of the confining potentials~(\ref{confpot}) and are, in general, rather difficult to compute due to their complex functional dependence on $\epsilon_{\sigma,n,k}$, $t_{n',k';n,k}$, and  $\phi_{\sigma,n,k}({\bm r})$. 

The analysis greatly simplifies in the {\em weak tunneling} regime where $\Delta_{\rm SAS}$ can be treated as a small perturbation compared to all other energy scales in the problem by employing the Born approximation. Consider, for instance, the event corresponding to elastic scattering ($\epsilon_{{\rm B},n,k} = \epsilon_{{\rm T},n',k'}$) from the unperturbed energy eigenstate $\phi_{{\rm B},n,k}({\bm r})$ (which describes an electron in the $n$-th Landau level, propagating with momentum $\hbar k$ along the $x$-axis on the B layer) to the state $\phi_{{\rm T},n',k'}({\bm r})$. Elementary manipulations of the ordinary equations of scattering theory~\cite{sakurai} in the Born approximation yield the following expression for the associated probability amplitude:
  \begin{eqnarray}
  &&S[\phi_{{\rm B},n,k} \rightarrow \phi_{{\rm T},n',k'}] \simeq (- 2\pi i) \nonumber\\
  &&\qquad \quad \times \left[-\frac{\Delta_{\rm SAS}}{2}\int d^2{\bm r}\;
  \frac{\phi^*_{{\rm T}, n', k'}({\bm r})}{\sqrt{\hbar v_{{\rm T},n',k'}   }} \frac{\phi_{{\rm B}, n, k}({\bm r})}{\sqrt{\hbar v_{{\rm B},n,k}     }}\right] 
  \nonumber \\
  &&\qquad \quad  =  \frac{i \Delta_{\rm SAS}}{2\hbar} \frac{e^{- i k k' \ell_B^2}}{\sqrt{v_{{\rm T},n',k'} v_{{\rm B},n,k} }} \; {\cal F}_{n',n}(k',k)~,   
 \label{scatte}
  \end{eqnarray}
where
  \begin{eqnarray} 
  v_{\sigma,n,k}= \frac{1}{\hbar} \frac{\partial \epsilon_{{\sigma}, n,k,}}{\partial k} \;, 
  \end{eqnarray}
is the group velocity of the mode $\phi_{\sigma,n,k}({\bm r})$ and where we have used Eq.~(\ref{tunn1}) to express the result in terms of the form factor (\ref{FormFactor}). Due to the perturbative nature of Eq.~(\ref{scatte}), its validity is restricted only to those cases where the modulus of $S[\phi_{{\rm B},n,k} \rightarrow \phi_{{\rm T},n',k'}]$ is small (we shall provide momentarily a more precise statement on this). 
Going beyond this regime is typically extremely challenging. 
 
Surprisingly, though, our scattering problem admits an explicit analytical solution in the special---but typically valid---scenario of \emph{smooth confinement}. In this limit the change of the confining potentials~(\ref{confpot}) over a magnetic length is assumed to be negligible with respect to the cyclotron gap, {\it i.e.} 
  \begin{eqnarray}\label{smooth}
  \ell_B \left| \frac{\partial V_{\sigma}(u_\sigma)}{\partial u_\sigma}\right| \ll \hbar \omega_{\rm c}~. 
  \end{eqnarray} 
This assumption well describes the physics of  edge states defined by electrostatic gates and has two main consequences, both  extremely useful in simplifying the scattering problem. 
Specifically,
\begin{itemize}
\item[{\em i)}] it implies a tunneling Hamiltonian~(\ref{coupling1}) where modes $\hat{c}_{\sigma,n,k}$,  $\hat{c}_{\sigma',n',k'}$  with $n\neq n'$
are effectively decoupled; smooth confining potentials obeying~(\ref{smooth}) do not mix edge beams living in the two layers and corresponding to different Landau-level indices.

\item[{\em ii)}] under proper conditions, it allows  the linearization of the dispersion relation of the unperturbed energy eigenvalues 
$\epsilon_{\sigma, n,k}$.
\end{itemize}
These  properties   originate from the fact that Eq.~(\ref{smooth}) permits to well approximate  the transverse wavefunctions $\chi_{\sigma, n,k}$ entering Eqs.~(\ref{Eq:WFB}) with the bulk Landau level eigenfunctions~\cite{Giuliani_and_Vignale}, {\it i.e.} with (properly translated) eigenstates of a harmonic oscillator of frequency $\omega_{\rm c}$ and zero-point fluctuation $\ell_B$, 
  \begin{eqnarray}
  \chi_{{\rm B}, n,k}(y)\simeq \ell_B^{-1/2}\varphi_n((y-k\ell_B^2)/\ell_B)\;, \nonumber \\
  \chi_{{\rm T}, n,k}(x)\simeq \ell_B^{-1/2}\varphi_n((x+k\ell_B^2)/\ell_B)\;, \label{transverselinAPP}
  \end{eqnarray}
where  $\varphi_n(\xi)=\frac{ \pi^{-1/4}}{\sqrt{2^n n!}}\;e^{-\xi^2/2}H_n(\xi)$, $H_n(\xi)$ being the $n$-th  Hermite polynomial. 
At the level of the unperturbed energy eigenvalues of the system this implies that one can write them as  a harmonic contribution plus a correction associated with the confinement potential, {\it i.e.} 
  \begin{eqnarray} \label{epsapp} 
  \epsilon_{\sigma,n,k} \simeq \hbar \omega_{\rm c}(n+1/2) + V_\sigma(k\ell_B^2)~,
  \end{eqnarray} 
which can then be turned into a linear expression in $k$ as required by property {\em ii)}  by properly limiting the interval of $k$  which enters the problem [see the following for details]. Property {\em i)} instead follows by observing that within  the approximation~(\ref{transverselinAPP})
the form factor (\ref{FormFactor}) of the system becomes   diagonal in $n$ and independent from the momenta $k$ and $k'$, {\it i.e.} 
  \begin{eqnarray} \label{eq:orthogonal}
  {\cal F}_{n',n}&\simeq & {\ell_B}  \int d\xi d\xi' \; e^{-i \xi\xi'}\; \varphi^*_{n'}(\xi) \varphi_n(\xi')\nonumber \\ \label{form1APP}
  &=&  \sqrt{2\pi}\; \ell_B\; i^{n'} \; \langle n'|n\rangle = \sqrt{2\pi}\; \ell_B\; i^{n}\;  \delta_{n',n}~.
  \end{eqnarray} 
For future reference, notice that, in writing the second identity, the operator algebra of the harmonic oscillator~\cite{Glauber} has been adopted to express ${\cal F}_{n',n}$ in terms of the system Fock states $|n\rangle$ and $|n'\rangle$. Eq.~(\ref{eq:orthogonal}) implies that the Landau-level index $n$ retains its validity as approximate quantum number even in the presence of tunneling. Hence we can drop the summation over $n$ in Eq.~(\ref{eigenfuntionint}) looking for eigenmodes of the form 
  \begin{eqnarray}
  \hat{\gamma}_n(\epsilon) &=& \sum_{\sigma = {\rm T}, {\rm B}}  \int dk \; (\alpha_{\sigma, n,k}^{(\epsilon)})^* \; \hat{c}_{\sigma,n,k} \;,\\
  {\psi}^{(\epsilon)}_{\sigma,n}({\bm r})&=&  \int dk \; \alpha_{\sigma, n,k}^{(\epsilon)} \;  \phi_{{\sigma},n,k,}({\bm r})\;, \label{eigenfuntionint111}
  \end{eqnarray}
where now $\alpha_{\sigma, n,k}^{(\epsilon)}$ solve the following simplified  system of coupled equations:
  \begin{eqnarray} \nonumber 
  (\epsilon_{{\rm B}, n,k}-\epsilon) \alpha_{{\rm B}, n,k}^{(\epsilon)} &=& \tfrac{(-i)^n \Delta_{\rm SAS} \ell_B}{2\sqrt{2\pi}}   \int {dk'}
  \; e^{i kk' \ell_B^2} \;\alpha_{{\rm T}, n,k'}^{(\epsilon)} \;,  \\\label{eigenEqintSM}\\
  (\epsilon_{{\rm T}, n, k}-\epsilon)  \alpha_{{\rm T}, n,k}^{(\epsilon)} &=& \tfrac{i^n \Delta_{\rm SAS} \ell_B}{2\sqrt{2\pi}}  \int dk'   \; e^{-i kk'   
  \ell_B^2} 
  \;\alpha_{{\rm B}, n,k'}^{(\epsilon)} \label{fundeq}  \;.  \nonumber 
  \end{eqnarray}

The smooth confinement condition~(\ref{smooth}) turns out to be useful also to better clarify  the limit of validity of the weak tunneling expression~(\ref{scatte}). Indeed, let us focus on the special case in which the two edge states involved in the scattering process belong to the lowest Landau levels of their respective layers ({\it i.e.} $n=n'=0$) and propagate with the same momentum $\hbar k$ under smooth confinement conditions. According to~(\ref{transverselinAPP}), in this case  the transverse wavefunctions $\chi_{\sigma, 0,k}$  can be approximated  by  Gaussian amplitude distributions with variance $\ell_B$. Their overlap vanishes in an area of radius $\sim \ell_B$ around the geometric crossing point of the associated classical skipping orbits, yielding a form factor proportional to $\ell_B$---see Eq.~(\ref{form1APP})---and a corresponding a scattering amplitude~(\ref{scatte}) with a  modulus that scales as  ${\Delta_{\rm SAS}\ell_B}/({\hbar v_{\rm F}})$. Here,  $v_{\rm F}$ is the associated group velocity of the two modes which we assume to be identical. 
The  condition for the validity of the perturbative approach can thus  be casted in terms of the following inequality:
  \begin{equation}\label{Eq:validityBorn}
\frac{\ell_B}{v_{\rm F}} \ll \frac{\hbar}{\Delta_{\rm SAS}}~.
  \end{equation}
This admits a simple physical interpretation in terms of  the ratio between the time $\ell_B/v_{\rm F}$ spent by an electron crossing the active tunneling region of size $\ell_B$ at a speed $v_{\rm F}$, and the time $\hbar/\Delta_{\rm SAS}$ that is necessary to tunnel from one layer to the other one. 

\subsection{A specific example: linear confinement}\label{sec:linear}

In this Section we illustrate more explicitly the notion of ``smooth confinement" by discussing a specific example, 
which is amenable to a fully-analytical treatment: the case of linear confinement~\cite{girvin}. 

Let us assume that  the potentials~(\ref{confpot})  have a linear dependence upon the transverse coordinate $u_\sigma$ in each layer:
  \begin{eqnarray}\label{eq:linearconfinement}
  V_{\rm B}(y)=eEy~, \qquad \qquad V_{\rm T}(x)=-eEx~,
  \end{eqnarray}
$E$ being the intensity of an applied uniform electric field.

With this choice, the unperturbed eigenenergies of the system acquire a linear dispersion on momentum $\hbar k$,
  \begin{eqnarray}\label{LINDISP}
  \epsilon_{\sigma,n,k}&=&\hbar\omega_{\rm c}(n+1/2) + \hbar v_{\rm F} k  - \frac{1}{2}m^* v^2_{\rm F}~, 
    \end{eqnarray} 
  characterized by a constant group velocity
  \begin{equation}\label{eq:groupvelocityconstant}
  v_{\sigma, n,k} = v_{\rm F}  \equiv \frac{e E \ell_B^2}{\hbar} = c \frac{E}{B}~.
 \end{equation}
Note that $v_{\rm F}$ coincides with the classical expression for the drift velocity in crossed uniform electric and magnetic field.
 
\begin{figure}[t]
\begin{center}
\includegraphics[width=0.4\textwidth]{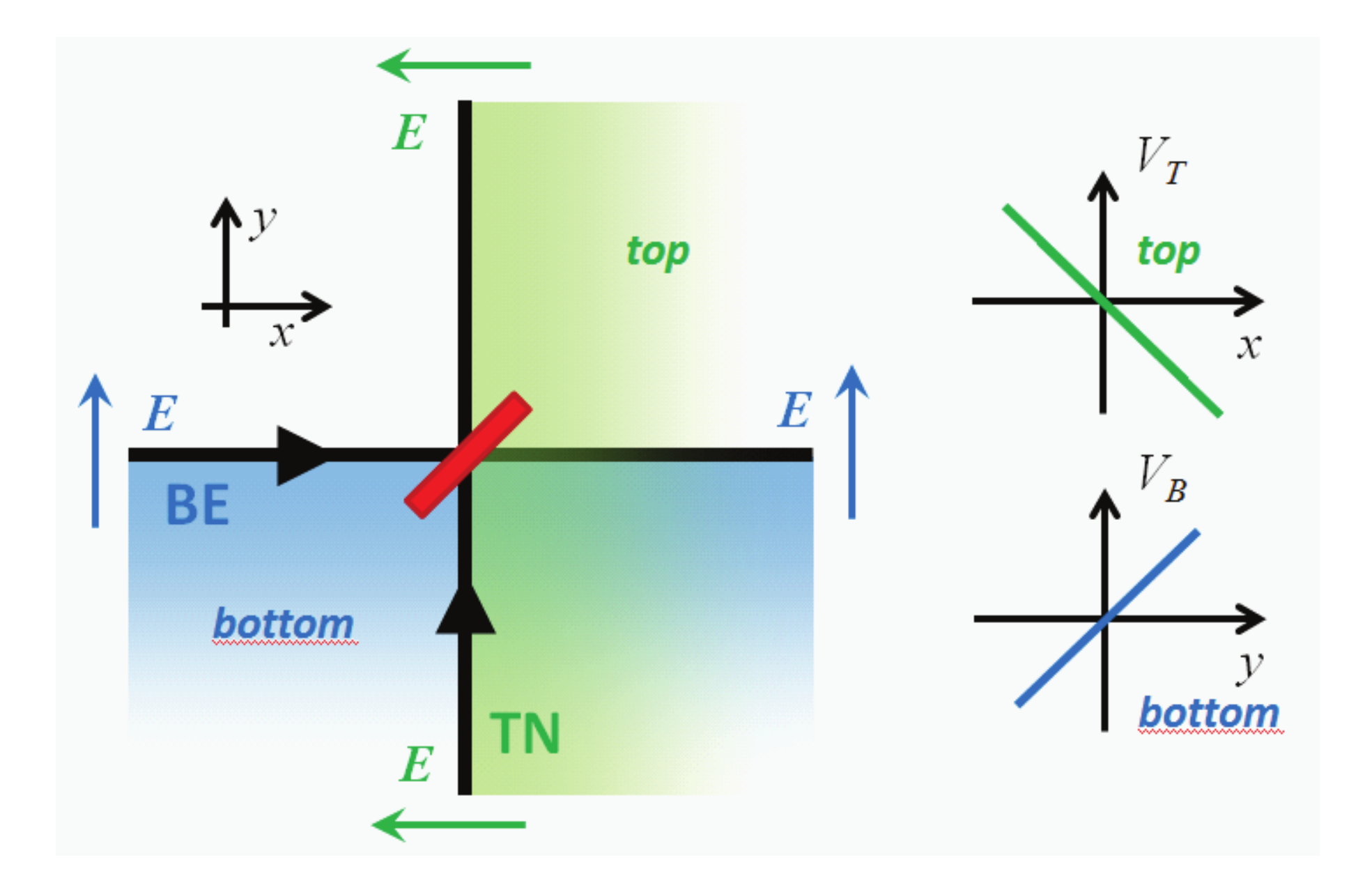}
\caption{(Color online) Interlayer edge mixing in the limit of a linear confinement potential, which obeys the smooth confinement condition (\ref{smooth}). Edge modes propagate west to east in the bottom layer and south to north in the top layer. Inter-layer tunneling is only active in a region of size $\approx\ell_B$ around the geometric crossing point of the edge modes. The mathematical analysis of the resulting scattering model leads to an ideal beam-splitter, closely resembling a semi-transparent mirror, with a transparency that depends monotonically 
on the strength of inter-layer tunneling $\Delta_{\rm SAS}$. The dependence of the confining potentials $V_\sigma(u_\sigma)$ on the transverse coordinate $u_\sigma$ is illustrated in the right panels.\label{fig:two}}
\end{center}
\end{figure}

Furthermore, the transverse wavefunctions $\chi_{\sigma,n,k}$ can be easily written in terms of properly translated harmonic-oscillator eigenstates:
  \begin{eqnarray}
  \chi_{{\rm B}, n,k}(y)=\ell_B^{-1/2}\varphi_n((y-k\ell_B^2)/\ell_B + \xi_0)\;, \nonumber \\
  \chi_{{\rm T}, n,k}(x)=\ell_B^{-1/2}\varphi_n((x+k\ell_B^2)/\ell_B - \xi_0)\;, \label{transverselin}
  \end{eqnarray}
where we have introduced the dimensionless parameter
  \begin{eqnarray}
  \xi_0 \equiv  \frac{ e E \ell_B}{\hbar \omega_{\rm c}}~.
  \end{eqnarray} 
Such functions describe chiral edges modes which, for $k>0$, propagate from west to east in the B layer, and from south to north in the T layer (see Fig.~\ref{fig:two}). The corresponding  form factor (\ref{FormFactor}) is independent of $k$ and $k'$ and  equal to 
 \begin{eqnarray} \label{eq:formfactorlinearpot}
 {\cal F}_{n',n}&=& {\ell_B}  \int d\xi d\xi' \; e^{-i (\xi+ \xi_0)(\xi'-\xi_0)}\; \varphi^*_{n'}(\xi) \varphi_n(\xi') \nonumber \\ \label{form1} 
 &=&  \sqrt{2\pi}\; \ell_B\; i^{n'} \;e^{{i}\xi_0^2/2}\langle n'|D\left(ie^{i\pi/4}\xi_0\right)|n\rangle~.
\end{eqnarray} 
Here, as in Eq.~(\ref{form1APP}), we have used the operator algebra of the harmonic oscillator~\cite{Glauber} to find the result expressed by the last equality. 
The displacement operator $D(\alpha)$ is defined by~\cite{Glauber}
\begin{equation}
D(\alpha) = \exp{(\alpha {\hat a}^\dagger - \alpha^* {\hat a})}~,
\end{equation}
where ${\hat a}$ (${\hat a}^\dagger$) is the harmonic-oscillator destruction (creation) operator.

The parameter $\xi_0$ gauges the smoothness of the linear potential with respect to the cyclotron gap. The criterion $\xi_0 \ll 1$ 
is indeed the smooth-confinement condition for the case of the linear-confinement model (\ref{eq:linearconfinement}). (Notice in particular that in the limit $\xi_0 \to 0$ Eq.~(\ref{transverselin}) yields Eq.~(\ref{transverselinAPP}).  Equation (\ref{eq:formfactorlinearpot})  makes it explicit that for $\xi_0\to 0$, the coupling between different Landau levels decreases exponentially with their distance, with a leading term which is proportional to $\xi_0^{|n'-n|}$. In particular, 
Eq.~(\ref{eq:formfactorlinearpot}) reduces to the diagonal expression~(\ref{form1APP}) in the limit $\xi_0 \to 0$, implying that tunneling only couples edge beams with the same $n$ and allowing us to look for eigenmode solutions of the form~(\ref{eigenfuntionint111}). Exploiting this fact and Eq.~(\ref{LINDISP}), the corresponding eigenmode equation~(\ref{eigenEqintSM}) can be finally casted in the following form:
  \begin{eqnarray} \left\{ 
  \begin{array}{l}
  [k_{\rm F}^{(n)}- k] \; \alpha_{{\rm B}, n,k}^{(\epsilon)} =\frac{{\gamma}(-i)^n}{\sqrt{2\pi}}  \;\int {dk'} \; e^{ikk' \ell_B^2} \; \alpha_{{\rm T}, 
  n,k'}^{(\epsilon)} \;,  \\\\\
  \label{fundeq11} \\\
  [k_{\rm F}^{(n)} - k] \;  \alpha_{{\rm T}, n,k}^{(\epsilon)} =\frac{{\gamma}\; i^n}{\sqrt{2\pi}}  \;  \int {dk'} \; e^{-ikk' \ell_B^2} \;\alpha_{{\rm B},   
  n,k'}^{(\epsilon)} \;,  \\
  \end{array}
  \right.
  \end{eqnarray}
where we have introduced the  quantities,
  \begin{eqnarray}
  \gamma &\equiv& - \frac{\Delta_{\rm SAS}}{2} \frac{\ell_B}{\hbar v_{\rm F}} \;, \\
  k_{\rm F}^{(n)} &\equiv& \frac{\epsilon - \hbar \omega_{\rm c} (n+1/2) + m^* v^2_{\rm F}/2}{\hbar v_{\rm F}}~.
  \end{eqnarray} 

As discussed in the Appendix~\ref{appex}, Eq.~(\ref{fundeq11}) admits analytical solutions of the form 
  \begin{eqnarray}
  \alpha^{(\epsilon)}_{{\rm B},n,k} &=&e^{i k k_{\rm F}^{(n)}\ell_B^2} 
  \;  f^{(0)}_{{\rm B}}(k \ell_B-k_{\rm F}^{(n)} \ell_B)\; \ell_B \label{alpha1}\\
  \alpha^{(\epsilon)}_{{\rm T},n,k} &=& 
  - i^n \; e^{i k_{\rm F}^{(n)}(k_{\rm F}^{(n)} - k)\ell_B^2}\; f^{(0)}_{{\rm T}}(k \ell_B - k_{\rm F}^{(n)} \ell_B))\; \ell_B~,
  \nonumber \\ \label{alpha2}
  \end{eqnarray} 
with 
  \begin{eqnarray} 
  {f}_{\rm B}^{(0)}(\kappa)  &=&   -\gamma \left[ c \; \Theta(\kappa) + d  \; \Theta(-\kappa)  \right]\; |\kappa|^{i \gamma^2}/\kappa \;,\label{fb} \\
  {f}_{\rm T}^{(0)}(\kappa)  &=& -\gamma  \left[ a \; \Theta(-\kappa) + b \; \Theta(\kappa)  \right]\; |\kappa|^{-i \gamma^2}/\kappa \;.\label{ft}
  \end{eqnarray}
Here $\Theta(\kappa)$ is the Heaviside step function, while $a$,$b$, $c$ and $d$ are  complex coefficients  
fixed by the boundary conditions.
These impose the following linear relationships between $c, d$ and $a,b$: 
  \begin{eqnarray}
  c  &=& \frac{\gamma}{\sqrt{2\pi}} \; \Gamma(-i \gamma^2) \; \left[ a \; e^{- \pi \gamma^2/2} - b \; e^{\pi \gamma^2/2}\right]~,  \nonumber \\
  d  &=& \frac{\gamma}{\sqrt{2\pi}} \; \Gamma(-i \gamma^2) \; \left[ a \; e^{\pi \gamma^2/2} - b \; e^{-\pi \gamma^2/2}\right]~,\label{constraint}
  \end{eqnarray} 
 where $\Gamma(\cdots)$ is the Euler Gamma function. By replacing~(\ref{alpha1}) and (\ref{alpha2}) into (\ref{eigenfuntionint111}) we finally get the eigenfunctions.
For the sake of simplicity, we here report only those associated with the lowest Landau level ({\it i.e.} $n=0$).
Specifically, introducing the dimensionless parameter $\xi_{\rm F}= k_{\rm F}^{(0)} \ell_B$ and the variables $\delta x = x+\xi_{\rm F} \ell_B$, $\delta y =  y -\xi_{\rm F}\ell_B$, and 
$z =(\delta  y  + i \delta x)/\ell_B$, one has
  \begin{eqnarray}
  {\psi}^{(\epsilon)}_{{\rm B},0}({\bm r})&=& -{\gamma} \; 
  \phi_{{\rm B}, 0,k_{\rm F}^{(0)}}({\bm r}) \; e^{ i \xi_{\rm F}^2}\;  \label{solution1}  \\
  && \times
  \left[ c\;\Lambda_{+} (-z-\xi_0)- d \; \Lambda_{+} (z+\xi_0)\right] \;,\nonumber \\
  {\psi}^{(\epsilon)}_{{\rm T}, 0}({\bm r})&=&  -{\gamma} \;   \phi_{{\rm T}, 0,k_{\rm F}^{(0)}}({\bm r}) \label{solution2} \\
  && \times \left[ - a \; \Lambda_{-} (iz+\xi_0) + b \; \Lambda_{-} (-iz - \xi_0)\right] \;,\nonumber 
  \end{eqnarray}
where $\phi_{{\rm \sigma}, 0,k_{\rm F}^{(0)}}({\bm r})$  are the eigenfunctions~(\ref{Eq:WFB}) at $\Delta_{\rm SAS} = 0$ and 
 \begin{eqnarray}\label{Lambda}
 \Lambda_{{\pm}} (z)&\equiv& \int_0^\infty d\xi \; e^{ -\xi^2/2 - \xi z} \; \xi^{\pm i \gamma^2-1}
 \nonumber \\
 &=& \Gamma(\pm i \gamma^2) \; e^{z^2/4} \; D_{\mp i\gamma^2} (z)~,
 \end{eqnarray}
$D_\kappa(z)$ being the parabolic cylinder special function. Interestingly, mathematically similar solutions have been independently obtained in a very recent work~\cite{blackhole} analyzing the QH effect in a 2DEG subject to a potential $\propto xy$. 
 
For $|z|\gg 1$  they admit the following asymptotic expansion
  \begin{eqnarray}\label{LambdaAsymptotic}
  \Lambda_{{\pm}} (z)&\simeq & \Gamma(\pm i \gamma^2)  \; z^{\mp i \gamma^2}\;,
  \end{eqnarray} 
which can be used to study the asymptotic behavior of the solutions~(\ref{solution1}) and (\ref{solution2}). In particular, from this it follows that in the limit $|\delta x /\ell_B| \gg 1$, the B-layer component behaves as
  \begin{eqnarray}\label{simp1}
  {\psi}^{(\epsilon)}_{{\rm B},0}({\bm r})&\simeq & 
  {\sqrt{2\pi}} \; 
  \phi_{{\rm B}, 0, k_{\rm F}^{(0)}}({\bm r}) \; e^{ i \xi_{\rm F}^2}\; e^{-i \gamma^2 \ln|\delta x/\ell_B|} \nonumber \\
  &\times& \left[ a \;\Theta(\delta x)+b \;  \Theta(-\delta x) \right]~,
  \end{eqnarray}
where  the constraint (\ref{constraint}) was employed in simplifying the expression. Similarly, for  $|\delta y/\ell_B| \gg 1$ we have 
  \begin{eqnarray}\label{simp2}
  {\psi}^{(\epsilon)}_{{\rm T},0}({\bm r})&=& {\sqrt{2\pi}}\;  \phi_{{\rm T}, 0,k_{\rm F}^{(0)}}({\bm r}) \; e^{i \gamma^2 \ln|\delta y/\ell_B|} \nonumber \\
  &\times& \left[ c \;\Theta(\delta y)+d \; \Theta(-\delta y) \right]~.
  \end{eqnarray} 
 %

\begin{figure}[top]
\begin{center}
\includegraphics[width=0.45\textwidth]{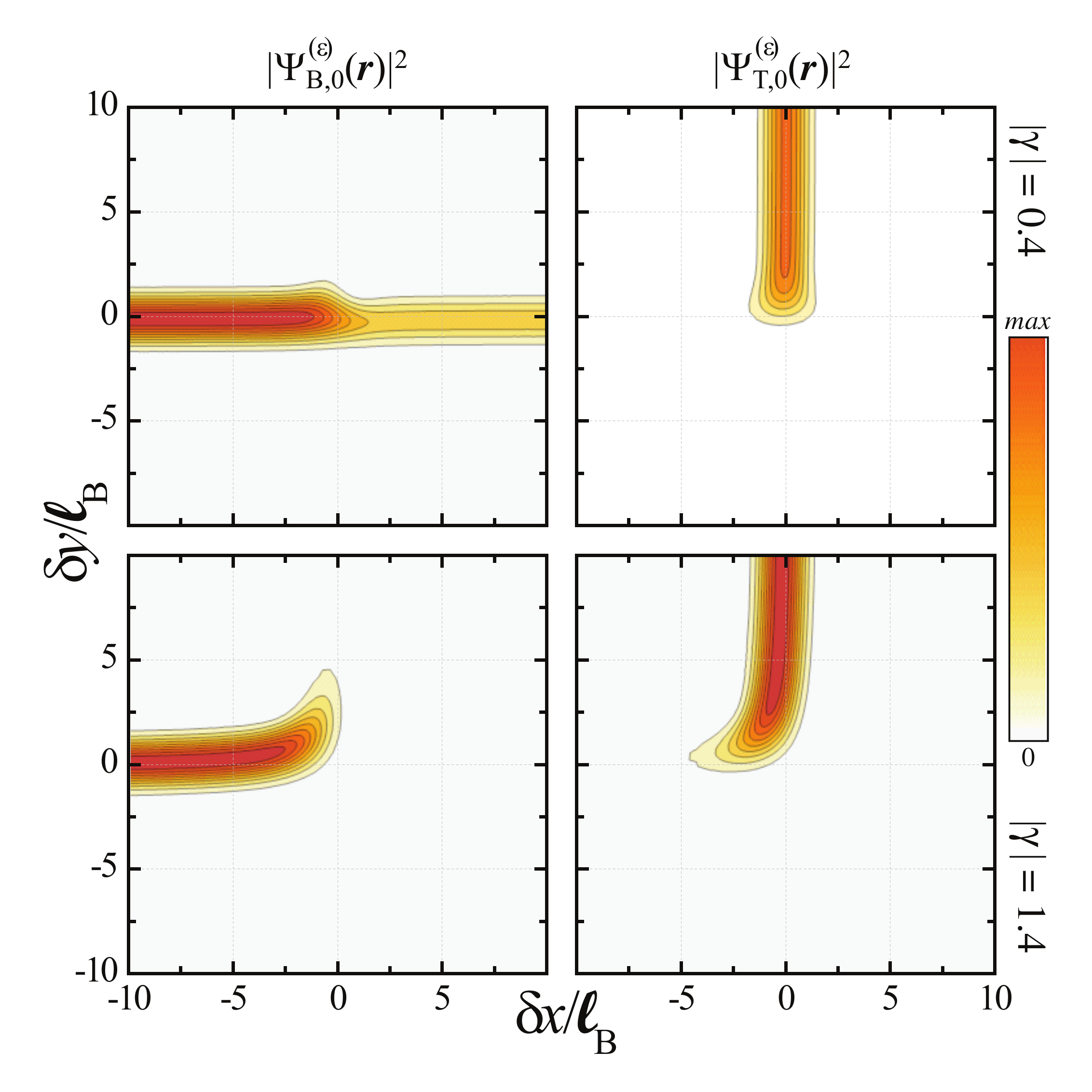}
\caption{(Color online) 2D color plots of $|\psi^{(\epsilon)}_{\sigma,0}({\bm r})|^2$ as obtained from Eqs.~(\ref{solution1}) and (\ref{solution2}). We remind the reader that these solutions refer to the case of linear confinement potentials, 
which obey the smooth-confinement condition (\ref{smooth}). The horizontal and vertical axes denote 
$\delta x/\ell_B$ and $\delta y/\ell_B$, respectively. The numerical results in this plot refer to the case in which an 
electron enters the system from the west in the B layer---{\it i.e.} $d=0$ and 
$c= - e^{-\pi \gamma^2/2 - i \xi^2_{\rm F}}/\gamma \Gamma(i\gamma^2)$. 
The left plots refer to the B-layer wavefunction $\psi^{(\epsilon)}_{{\rm B},0}({\bm r})$, while the right plots to the T-layer wavefunction 
$\psi^{(\epsilon)}_{{\rm T},0}({\bm r})$. The upper panels refer to $\gamma=-0.4$, while the lower panels to $\gamma=-1.4$. 
We have taken $\xi_0 = 0$ in Eqs.~(\ref{solution1}) and (\ref{solution2}), consistently with the smooth-confinement condition (\ref{smooth}). \label{fig:three}}
\end{center}
\end{figure}

\begin{figure}[top]
\begin{center}
\includegraphics[width=0.45\textwidth]{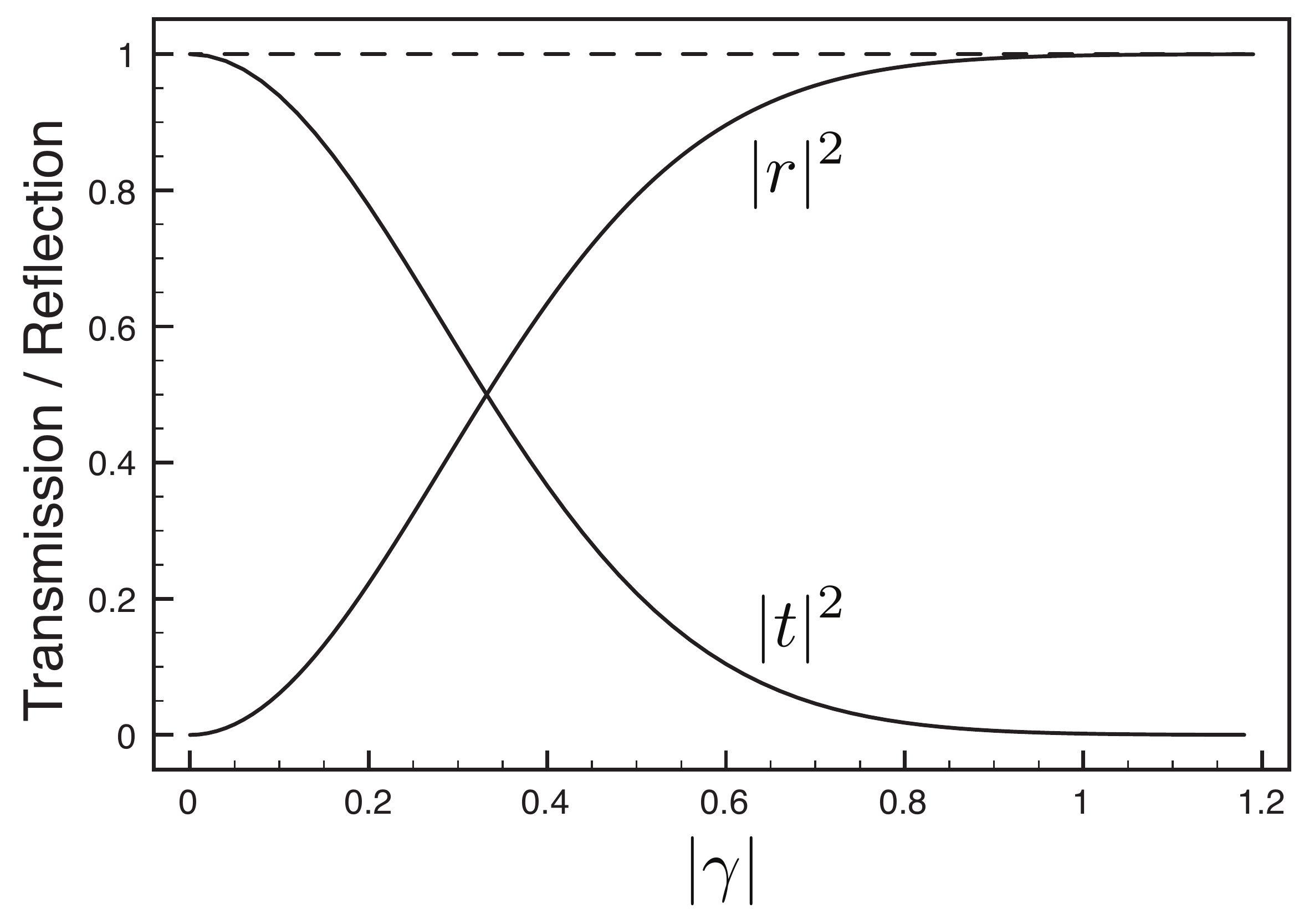}
\caption{The transmission $|t|^2$ and reflection $|r|^2$ probabilities of the linear confinement problem as obtained from Eqs.~(\ref{transm})-(\ref{erre}) are plotted as functions of $|\gamma|$. 
The dashed line represents $|r|^2+|t|^2 = 1$.\label{fig:four}}
\end{center}
\end{figure}

Equations~(\ref{simp1}) and (\ref{simp2}) define a set of plane-waves  impinging on/emerging from the crossing point 
  ${\bm r}_0 = (-\xi_{\rm F} \ell_B + \xi_0,\xi_{\rm F} \ell_B - \xi_0)$ of the unperturbed chiral edge modes---the logarithmic phase terms being irrelevant corrections when compared to the longitudinal
phase dependence of $\phi_{{\rm B}, 0, k_{\rm F}^{(0)}}({\bm r})$ and $\phi_{{\rm T}, 0, k_{\rm F}^{(0)}}({\bm r})$. 
For instance, setting $d=0$ and $c= - e^{-\pi \gamma^2/2 - i \xi^2_{\rm F}}/\gamma \Gamma(i\gamma^2)$ we obtain
\begin{eqnarray}
{\psi}^{(\epsilon)}_{{\rm B},0}({\bm r})&\simeq & 
 \left[t \;  \Theta(\delta x)+  \Theta(-\delta x) \right] \; \phi_{{\rm B}, 0, k_{\rm F}^{(0)}}({\bm r})\;,\nonumber \\
{\psi}^{(\epsilon)}_{{\rm T},0}({\bm r})&\simeq & \label{simp12}
 r \; \Theta(\delta y) \; \phi_{{\rm T}, 0, k_{\rm F}^{(0)}}({\bm r})~,
\end{eqnarray}
which describe a scattering event where an incoming wave from the west hand side of the B layer---represented
by the component $\psi_{\rm B}^{({\rm in})}({\bm r}) \equiv \Theta(-\delta x)  \phi_{{\rm B}, 0, k_{\rm F}^{(0)}}({\bm r})$---splits into a  transmitted wave 
$\psi_{\rm B}^{({\rm out})}({\bm r}) \equiv \Theta(\delta x)  \phi_{{\rm B}, 0, k_{\rm F}^{(0)}}({\bm r})$ propagating on the east hand side of the same layer,
and into a  deflected  wave $\psi_{\rm T}^{({\rm out})}({\bm r}) \equiv \Theta(\delta y)  \phi_{{\rm T}, 0, k_{\rm F}^{(0)}}({\bm r})$ 
propagating along the south-north direction in the T layer---see Fig.~\ref{fig:two}. The corresponding transmissivity $t$ and reflectivity $r$ are determined by the parameters
\begin{eqnarray}
t &\equiv& e^{-\pi \gamma^2} \;, \label{transm}\\
r &\equiv&-\frac{\sqrt{2\pi}}{\gamma} \frac{e^{-\pi \gamma^2/2}}{\Gamma(i\gamma^2)} \; e^{- i \xi_{\rm F}^2}~, \label{erre}
\end{eqnarray} 
which fulfill the normalization condition $|r|^2 + |t|^2 = 1$ thanks to the identity
 $|\Gamma(i\gamma)|^2 = {2\pi}~(e^{\pi \gamma^2} - e^{-\pi \gamma^2})^{-1}/\gamma^2$---see Fig.~\ref{fig:four}.
 
Notice, in particular, that in the weak tunneling $\gamma \to 0$ limit,  the reflectivity $r$ in Eq.~(\ref{erre}) reduces to the value 
  \begin{eqnarray} \label{BORN11}
 r \simeq -i \; e^{-i\xi_{\rm F}^2} \;  \sqrt{2\pi} \gamma = i \; e^{- i\xi_{\rm F}^2} \; \sqrt{\frac{\pi}{2}}\;  \frac{\ell_B\Delta_{\rm SAS}}{\hbar v_{\rm F}}~,
 \end{eqnarray} 
in perfect agreement with the result based on the Born approximation. The latter is obtained from Eq.~(\ref{scatte}) by taking 
 $\phi_{{\rm B}, 0, k_{\rm F}^{(0)}}({\bm r})$ and 
 $\phi_{{\rm T}, 0, k_{\rm F}^{(0)}}({\bm r})$   as input and output states.  

In a similar fashion, setting $b=0$ and $a=  e^{-\pi \gamma^2/2 }/\gamma \Gamma(-i\gamma^2)$, Eqs.~(\ref{simp1}) and (\ref{simp2})  can be used to describe the
 scattering event ``complementary" to~(\ref{simp12}) where  the incoming wave $\psi_{\rm T}^{({\rm in})}({\bm r}) \equiv \Theta(-\delta y)  \phi_{{\rm T}, 0, k_{\rm F}^{(0)}}({\bm r})$  
 is reaches ${\bm r}_0$ from the south-north direction in the T layer and gets partially deflected in the B layer. In this case we get 
\begin{eqnarray}\label{simp11}
{\psi}^{(\epsilon)}_{{\rm B},0}({\bm r})&\simeq & 
- r^*\;  \Theta(\delta x)\; \phi_{{\rm B}, 0, k_{\rm F}^{(0)}}({\bm r})  \nonumber \;,\\
{\psi}^{(\epsilon)}_{{\rm T},0}({\bm r})&\simeq & \left[ t^* \Theta(\delta y)+  \Theta(-\delta y) \right] \; 
\phi_{{\rm T}, 0, k_{\rm F}^{(0)}}({\bm r})~,
\end{eqnarray}
which, together with Eq.~(\ref{simp11}), ensure the unitarity of the mapping $[\psi_{\rm B}^{({\rm in})}({\bm r}),\psi_{\rm T}^{({\rm in})}({\bm r})] \rightarrow[\psi_{\rm B}^{({\rm out})}({\bm r}),\psi_{\rm T}^{({\rm out})}({\bm r})]$.

\section{Exact solution for general potentials in the smooth confinement limit}
\label{sect:exactsolution}

In this Section we address the case of confinement potentials which are smooth, {\it i.e.} they satisfy Eq.~(\ref{smooth}), but not necessarily linearly-dependent on the transverse coordinate $u_\sigma$, as in the previous Section. In this case we need to solve Eq.~(\ref{eigenEqintSM}) with $\epsilon_{\sigma,n,k}$ as in Eq.~(\ref{epsapp}). 

The idea we are going to exploit is to focus the attention on an interval of values of $k$ around which the dispersion relation can be linearized. Specifically, we assume that around point A in Fig.~\ref{fig:one} the following expansion holds
 \begin{eqnarray} \label{linappr}
\epsilon_{\sigma, n,k} \simeq \epsilon_{\sigma, n, k_{\rm F}} + (k - k_{\rm F}) \hbar v_{\rm F}  + \dots
\end{eqnarray}
for $k \in (k_{\rm F} - \Delta k/2, k_{\rm F} + \Delta k/2)$.
$k_{\rm F}$ is the Fermi momentum of the system associated with the energy $\epsilon$, {\it i.e.} 
$\epsilon_{\sigma, n, k_{\rm F}} = \epsilon$, while $v_{\rm F}$ is the group velocity of the modes,
\begin{eqnarray}
v_{\rm F} = \left. \frac{\partial \epsilon_{\sigma, n,k}}{\partial k} \right|_{k=k_{\rm F}} = 
\left. \frac{\partial V_{\sigma}(k \ell^2_B)}{\partial k} \right|_{k=k_{\rm F}}~,
\end{eqnarray}
which, for the sake of simplicity, we assume to be layer independent.
Notice, however,  that, in principle, for each $n$ one has 
specific values of $k_{\rm F}$, $\Delta k$, and $v_{\rm F}$. 

For this purpose, in analogy with the layer field operator $\hat\Psi_\sigma({\bm r})$ in Eq.~(\ref{expansion}), we now define the layer-resolved edge field
\begin{equation}\label{EdgeField}
\hat\xi_{\sigma,n}(r_\sigma)=\int \frac{dk}{\sqrt{2\pi}}~e^{i k r_\sigma}g_{k}~\hat{c}_{\sigma,n,k}~,
\end{equation}
with $r_{\rm B}=x$ ($r_{\rm T} = y$) for the B (T) layer and
\begin{equation}
g_{k}=\exp\left[-\frac{(k-k_{\rm F})^2}{2\Delta k^2}\right]~.
\end{equation}
The introduction of the kernel $g_{k}$ has two consequences: {\it i)} it forces $k$ to lie in the interval where~(\ref{linappr}) holds;  {\it ii)} it induces a coarse graining in the position resolution of the fields, which now obey fermionic anti-commutation rules only beyond a length scale $2\pi/\Delta k$,
\begin{eqnarray}
\{\hat\xi_{\sigma,n}(r_\sigma),\hat\xi^{\dag}_{\sigma',n'}(r'_{\sigma'})\} &=&\delta_{\sigma,\sigma'}\delta_{n,n'}e^{ik_{\rm F}(r_\sigma-r'_\sigma)}\nonumber\\
&\times&\delta_{\Delta k}(r_\sigma - r'_\sigma),
\end{eqnarray}
with $\delta_{\Delta k}(x)=\Delta k(2\sqrt{\pi})^{-1}\exp(-x^2\Delta k^2/4)$.

The edge field $\hat\xi_{\sigma,n}(r)$ in Eq.~(\ref{EdgeField}) can be equivalently expanded by using the field $\hat\gamma(\epsilon)$, which satisfies the eigenmode equation~(\ref{eigenmode}),
\begin{equation}
\hat\xi_{\sigma,n}(r)=\sum_\epsilon\; \xi_{\sigma,n}^{(\epsilon)}(r)\; \hat\gamma(\epsilon).
\end{equation}
By exploiting Eq.~(\ref{unitarymap}), which connects $\hat\gamma(\epsilon)$ to the unperturbed fields $\hat{c}_{\sigma,n,k}$, and which can be inverted as
\begin{equation}
\hat{c}_{\sigma,n,k}=\sum_\epsilon\; \alpha_{\sigma,n,k}^{(\epsilon)}\hat\gamma(\epsilon)~,
\end{equation}
we can identify, in complete analogy with the layer wavefunctions Eq.~(\ref{eigenfuntionint}),  the {\em coarse grained} edge wavefunctions,
\begin{eqnarray}
\xi^{(\epsilon)}_{\sigma,n}(r_\sigma) \equiv \int \frac{dk}{\sqrt{2\pi}} \; e^{i k r_\sigma} \; g_k\; \alpha^{(\epsilon)}_{\sigma,n,k}~.\label{waveappr}
\end{eqnarray} 

In the smooth-confinement approximation, Eq.~(\ref{smooth}), the complex amplitudes $\alpha^{(\epsilon)}_{\sigma,n,k}$ satisfy Eq.~(\ref{eigenEqintSM}). The introduction of the  distribution $g_k$ effectively linearizes the dispersion $\epsilon_{\sigma,n,k}$ for those $k$ which belong to the interval $(k_{\rm F} - \Delta k/2, k_{\rm F} + \Delta k/2)$. It then follows that the wavefunctions $\xi^{(\epsilon)}_{\sigma,n}(r_\sigma)$ satisfy the following equation
\begin{eqnarray}
(k_{\rm F}+i\partial_x)\xi^{(\epsilon)}_{{\rm B}, n}(x)=\frac{(-i)^n\gamma}{\sqrt{2\pi}}\int dy~\xi^{(\epsilon)}_{{\rm T}, n}(y)K(x,y),\nonumber\\\\
(k_{\rm F}+i\partial_y)\xi^{(\epsilon)}_{{\rm T}, n}(y)=\frac{i^n\gamma}{\sqrt{2\pi}}\int dx~\xi^{(\epsilon)}_{{\rm B}, n}(x)K^*(-y,-x),\nonumber
\end{eqnarray}
where the kernel $K(x,y)$ is obtained under the condition $\Delta k\ell_B\gg 1$ and reads 
\begin{equation}
K(x,y)=\frac{1}{\ell_B^2}\exp\left[-\frac{(y - k_{\rm F}\ell_B^2)^2}{2\Delta k^2\ell_B^2} + i\frac{xy}{\ell_B^2}\right]~.
\end{equation}
By further tightening the condition $\Delta k\ell_B\gg 1$ we can cast the system of equation for the wavefunctions $\xi_{\sigma,n}^{(\epsilon)}(r_\sigma)$ in the form
\begin{eqnarray}
(k_{\rm F}+i\partial_x)\xi^{(\epsilon)}_{{\rm B}, n}(x)&=&(-i)^n\gamma\; \tilde\xi^{(\epsilon)}_{{\rm T}, n}(-y/\ell_B^2)~,\\
(k_{\rm F}+i\partial_y)\xi^{(\epsilon)}_{{\rm T}, n}(y)&=&i^n\gamma\; \tilde\xi^{(\epsilon)}_{{\rm B}, n}(x/\ell_B^2)~.
\end{eqnarray}
The solutions of these equations are given by Eqs.~(\ref{fB-tilde},\ref{fT-tilde}), which, by neglecting the logarithmic phase $\exp(\pm i\gamma^2\ln|r_\sigma/\ell_B|)$ with respect to the linear increase, read 
\begin{eqnarray}
\xi^{(\epsilon)}_{{\rm B}, n}(x)&\simeq&e^{ik_{\rm F} x}e^{i(k_{\rm F}\ell_B)^2}\left[a\; \Theta(x)+b\; \Theta(-x)\right],\label{EdgeFieldsAsymptoticB}\\ \label{EdgeFieldsAsymptoticT}
\xi^{(\epsilon)}_{{\rm T}, n}(y)&\simeq& e^{ik_{\rm F} y}\left[c\; \Theta(y)+d\; \Theta(-y)\right].
\end{eqnarray}
The coarse-graining procedure has averaged out all the details below a scale $2\pi/\Delta k$, which is much smaller than the magnetic length $\ell_B$, which in turn is the smallest length scale in the problem. Eqs.~(\ref{EdgeFieldsAsymptoticB},\ref{EdgeFieldsAsymptoticT}) are in complete analogy with Eqs.~(\ref{simp1}, \ref{simp2}). We can then extend the results for the reflection and transmission amplitudes Eqs.~(\ref{transm}, \ref{erre}) derived in the special case of linear confinement to the general case of smooth confinement.  

\section{Interferometer scattering matrix}
\label{Sec:ScatteringMatrix}

The full response of the interferometer can be studied in the Landauer-B\"uttiker formalism~\cite{Buttiker,Datta} by means of the scattering matrix $S$.  In the previous Sections we have focused the attention on the mixing between two particular edge states, ``BE", which propagates 
from west to east in the bottom layer, and  ``TN", which propagates from south to north in the top layer. These two modes cross at position $A$ in Fig.~\ref{fig:one}. In particular, we calculated the corresponding transmission amplitude $t$, Eq.~(\ref{transm}), and the reflection amplitude $r$, Eq.~(\ref{erre}). 

The scattering at the other crossing points B, C, and D can be characterized in a completely analogue way and we can write the scattering matrix $s_i$ characterizing each beam-splitter $i= {\rm A}, {\rm B}, {\rm C}$, and ${\rm D}$ as
\begin{equation}\label{Eq:SeffBS}
s_i=\left[\begin{array}{cc}
t_i & r_i\\
-r^*_i & t^*_i\end{array}\right],
\end{equation}
with $t_i$ and $r_i$ given by $t$ and $r$, Eq.~(\ref{transm}, \ref{erre}), respectively, for nominally equal beam-splitters. 

The full scattering matrix $S$ that characterizes the interferometer response can now be obtained through a concatenation procedure \cite{Datta}, staring from the scattering matrix $s_i$ of the elementary beam-splitter. For instance, the amplitude $S_{21}$ for scattering from contact $1$ to contact $2$---see Fig.~\ref{fig:one}---can be obtained as a geometric series that sums all the possible paths an electron can take before exiting from $2$, and for nominally equal beam-splitters takes the form 
\begin{eqnarray}
S_{21}=r-e^{i\varphi_{\rm dyn}}\frac{|t|^2(r^*)^3}{1-(r^*)^4e^{i\varphi_{\rm dyn}}}~,
\end{eqnarray}
where the phase $\varphi_{\rm dyn}=4k_{\rm F}W$ is the dynamical phase acquired around the perimeter of the square defined by the corners $i= {\rm A}, {\rm B}, {\rm C}$, and ${\rm D}$ in Fig.~\ref{fig:one}.
The transmission probability $T_{21} = |S_{21}|^2$ can be written as
\begin{eqnarray}
T_{21}  = R\left|1-e^{i\varphi}\frac{R(1-R)}{1-R^2e^{i\varphi}}\right|^2,
\end{eqnarray} 
where $R=|r|^2$ and $\varphi=\varphi_{\rm dyn} +4(k_{\rm F}\ell_B)^2$, up to an offset given by the argument of $\Gamma(-i\gamma^2)$.
We notice that the phase $4(k_{\rm F}\ell_B)^2$ is equal to the Aharonov-Bohm phase $2\pi\Phi/\Phi_0$ ($\Phi_0=hc/e$ being the flux quantum), where $\Phi$ is the flux through the area of the square ABCD defined by the edge state crossing points of coordinates $x_i=\pm k_{\rm F}\ell_B^2$ and $y_i=\pm k_{\rm F}\ell_B^2$.

$T_{21}$ is zero for $R=0$ (no tunneling) and one for $R=1$, (all the current injected in contact 1 is totally drained at the first beam-splitter at point A). Analogously, one can obtain all the other transmission and reflection amplitudes.

\section{Discussion and conclusions}
\label{sect:conclusions}

We have proposed a strategy for the implementation of electron interferometry which is based on quantum Hall bilayers. Our approach 
exploits uniform inter-layer tunneling to mix edge modes. While tunneling between the two layers is present in the whole region where 
top and bottom Hall bars overlap (see Fig.~\ref{fig:one}), the quasi one-dimensional character of edge states implies that inter-edge tunneling is active only in a region of linear size of order of the magnetic length around the geometrical crossing points of edge states belonging to different layers. 

We have demonstrated that the scattering problem associated with the quantum Hall bilayer depicted in Fig.~\ref{fig:one} can be solved 
{\em exactly} provided that the confining potentials defining the two Hall bars obey the ``smooth confinement" condition (\ref{smooth}). 
The analysis of this Article relies on a single-particle picture and therefore neglects inter-layer electron-electron interactions. Fascinating many-body phenomena including spontaneous inter-layer coherence and Coulomb drag effects are expected to occur in QH bilayers~\cite{manybody}. These effects are particularly spectacular in the regime in which $\Delta_{\rm SAS}$ is a negligible energy scale. In this Article we have analyzed the opposite regime. Electron-electron interactions in the regime $\Delta_{\rm SAS} \ll e^2/(\epsilon d)$, where $\epsilon$ is a material parameter ($\epsilon \sim 13$ for GaAs), might have a non-trivial impact on the behavior of the proposed beam splitter. We hope to tackle this intriguing regime in a forthcoming publication.

Turning to interferometric implementations, the key advantage of the proposed setup is that it bypasses the use of quantum point contacts to mix edge modes. ``Non-simply connected circuits" are not necessary for the achievement of our interferometer. This is expected to allow the design of smaller and more advanced coherent circuits in the quantum Hall regime. 

Our quantum Hall bilayer beam splitter is expected to display a behavior that closely mimics the one of a semi-transparent mirror in conventional optics. One drawback of our proposal is linked to the fact that in bilayers realized in semiconductor heterojunctions the magnitude of the interlayer tunneling $\Delta_{\rm SAS}$ is mostly determined by the heterostucture design (and thus fixed by the growth procedure). On the other hand, $\Delta_{\rm SAS}$ can be tuned by an in-plane magnetic field~\cite{Hu92}, which, for sufficiently thin samples, does not introduce additional orbital effects. This implies a certain degree of tunability of the beam splitter transmittivity. Another degree of tunability, which we have not exploited  in the present Article, is offered by the relative angle $\theta$ between edge modes living in different layers (in this Article we have analyzed only $\theta = \pi/2$). 

A very interesting perspective is also constituted by the possibility to implement a beam splitter in the fractional quantum Hall regime, which has been historically hard. While it is not obvious at all how to generalize the current theory to such regime, a similar phenomenology might be expected and might shed new light on transport phenomena involving fractional edge modes. Last but not least, we would like to mention that coherent beam splitters may also be realized in vertical heterostructures~\cite{verticalheterostructures} comprising graphene as well as other two-dimensional crystals~\cite{novoselov_PNAS_2005,wang_naturenano_2012} such as MoS$_2$, h-BN, etc. The theoretical complication posed by these structures, however, is highly non-trivial since a tunneling Hamiltonian as simple as the one in Eq.~(\ref{INT}) does not apply. 
Inter-layer transport in these systems is very interesting~\cite{twistedlayers} and currently far from being completely understood.

\section*{Acknowledments}
V.G. wishes to thank A.S. Holevo for discussions and comments. 
This work was supported by MIUR through the programs 
``FIRB IDEAS" - Project ESQUI (Grant No. RBID08B3FM), ``FIRB - Futuro in Ricerca 2010" - Project PLASMOGRAPH (Grant No. RBFR10M5BT) and by the EU FP7 Programme under Grant Agreements No. 234970-NANOCTM.

\appendix

\section{Derivation of the solutions~(\ref{alpha1}) and~(\ref{alpha2})} \label{appex}

A convenient way to write Eq.~(\ref{fundeq11}) is by  introducing the  dimensionless variable $\kappa=k\ell_B$  and the complex functions  $f_{\rm B}(\kappa)$ and $f_{\rm T}(\kappa)$ defined implicitly by the expressions
  \begin{eqnarray}
  \alpha^{(\epsilon)}_{{\rm B},n,k} &=&
  \;  f_{{\rm B}}(k \ell_B) \; \ell_B\;,\\
  \alpha^{(\epsilon)}_{{\rm T},n,k} &=& (-i)^n 
  f_{{\rm T}}(k \ell_B) \; \ell_B \;. \label{alpha01}
  \end{eqnarray} 
Via Fourier transform Eq.~(\ref{fundeq11}) can then
be expressed as 
  \begin{eqnarray} \left\{ 
  \begin{array}{l}
  (\kappa_{\rm F}+ i \frac{d}{d \kappa}) \; \tilde{f}_{\rm B}(\kappa) =
  \gamma \; {f}_{\rm T}(-\kappa) \;,  \\  \label{equation11} \\
  (\kappa_{\rm F}+\kappa) \;  f_{\rm T}(-\kappa)  =
  \gamma \; \tilde{f}_{\rm B}(\kappa)\;,
  \end{array}
  \right.
  \label{newsyst}
  \end{eqnarray} 
or, equivalently, as 
  \begin{eqnarray} 
  \left\{ \begin{array}{l}
  (\kappa_{\rm F}- \kappa) \; f_{\rm B}(\kappa) =
  \gamma \; \tilde{f}_{\rm T}(\kappa) \;,  \\ \label{fff1} \\
  (\kappa_{\rm F}+ i  \frac{d}{d \kappa} ) \;  \tilde{f}_{\rm T}(\kappa)  =
  \gamma \; {f}_{\rm B}(\kappa)\;,
  \end{array} \right.
  \end{eqnarray} 
where $\kappa_{\rm F}$  stands for the dimensionless parameter $\kappa_{\rm F} \equiv k_{\rm F}^{(n)}\ell_B$, while we used the symbol $\tilde{f}(\kappa)$ to  represent the Fourier transform of the function $f(\kappa)$, {\it i.e.} 
  \begin{eqnarray}
  \tilde{f}(\kappa) = \frac{1}{\sqrt{2\pi}} \int d\kappa' \; e^{i \kappa'\kappa} \; f(\kappa')\;.
  \end{eqnarray}
By a close inspection of these equations, one notices that they fulfill certain symmetries. In particular  they are  invariant in form 
by replacing $f_{\rm B}(\kappa)$  with the complex conjugate of $f_{\rm T}(\kappa)$ and vice-versa, {\it i.e.} 
  \begin{eqnarray}
  (f_{\rm B}(\kappa), f_{\rm T}(\kappa)) \longrightarrow (f_{\rm T}^*(\kappa), f_{\rm B}^*(\kappa)) \;.
  \end{eqnarray}
Furthermore it is easy to check that the solutions ${f}_{\rm B}(\kappa)$, $f_{\rm T}(\kappa)$ for the case $\kappa_{\rm F}\neq 0$ can be written as 
  \begin{eqnarray}
  \tilde{f}_{\rm B}(\kappa) &=&e^{i \kappa_{\rm F}(\kappa+ \kappa_{\rm F})}  \tilde{f}_{\rm B}^{(0)}(\kappa+\kappa_{\rm F})\;,  \label{fB-tilde}\\ 
  {f}_{\rm T}(\kappa) &=&e^{i \kappa_{\rm F}(\kappa_{\rm F}-\kappa)} {f}_{\rm T}^{(0)}(\kappa-\kappa_{\rm F})\;,  \label{qui}
  \end{eqnarray}
or, equivalently,
  \begin{eqnarray}
  {f}_{\rm B}(\kappa) &=&e^{i \kappa_{\rm F}\kappa}  {f}_{\rm B}^{(0)}(\kappa-\kappa_{\rm F})\;,  \\ 
  \tilde{f}_{\rm T}(\kappa) &=&e^{i \kappa_{\rm F}\kappa} \tilde{f}_{\rm T}^{(0)}(\kappa-\kappa_{\rm F})\;, \label{fT-tilde}
  \end{eqnarray}
with ${f}_{\rm B}^{(0)}(\kappa), f_{\rm T}^{(0)}(\kappa)$ solving  Eq.~(\ref{newsyst}) [or~(\ref{fff1})] for $\kappa_{\rm F}=0$. Explicit expressions for the latter can then be obtained by observing that $\tilde{f}_{\rm B}^{(0)}(\kappa)$ fulfills the following (complex) Euler differential equation, 
  \begin{eqnarray}
  \kappa \;  \frac{d}{d \kappa} \; \tilde{f}_{\rm B}^{(0)}(\kappa) =
  -i \; \gamma^2 \;\tilde{f}_{\rm B}^{(0)}(\kappa)\;,
  \label{newsyst1}
  \end{eqnarray} 
[this can be verified via elementary algebraic manipulation of Eq.~(\ref{equation11})]. Upon integration, we thus obtain 
  \begin{eqnarray} 
  \tilde{f}_{\rm B}^{(0)}(\kappa)  &=&  \left[ a \; \Theta(\kappa) + b  \; \Theta(-\kappa)  \right]\; |\kappa|^{-i \gamma^2}\;,\label{sol11} 
  \end{eqnarray}
and, via the second line of Eq.~(\ref{newsyst}),  
  \begin{eqnarray} 
  {f}_{\rm T}^{(0)}(\kappa)  = -\frac{\gamma}{\kappa}  \left[ a \; \Theta(-\kappa) + b \; \Theta(\kappa)  \right]\; |\kappa|^{-i \gamma^2} \;,\label{sol12}
  \end{eqnarray}
which, together with~(\ref{qui}), gives us~(\ref{alpha2}) when replaced into~(\ref{alpha01}). In these expressions $\Theta(\kappa)$ is the Heaviside step function, while $a$, $b$ are complex coefficients that are fixed by  the boundary conditions fulfilled by $\tilde{f}_{\rm B}^{(0)}(\kappa)$ and ${f}_{\rm T}^{(0)}(\kappa)$. One might notice that the solutions~(\ref{sol11}) and (\ref{sol12})  are well defined everywhere but for $\kappa=0$ where they are unstable due to the fast oscillating term  $|\kappa|^{-i\gamma^2}$ and, in the case of ${f}_{\rm T}^{(0)}(\kappa)$, due to the presence of the factor $1/\kappa$ [notice however that the presence of ``stronger" discontinuities  ({\it e.g.} Dirac delta or differential Dirac delta contribution terms) can be excluded by a close  inspection of the differential equation~(\ref{newsyst1})]. These irregularities   cannot be avoided and need to be properly accounted for when producing the functions   $\tilde{f}_{\rm T}^{(0)}(\kappa)$  and ${f}_{\rm B}^{(0)}(\kappa)$ via Fourier (or inverse Fourier) transformation.
The correct prescription is obtained by adopting the following integral expressions
  \begin{eqnarray} 
  &&\int_0^{\infty} d\kappa'  \;  {[\kappa']^{i\gamma^2-1}} \; e^{i \kappa\kappa'} = \lim_{\delta \rightarrow 0^+} 
  \frac{\Gamma(i\gamma^2+\delta)}{(-i \kappa)^{i\gamma^2+\delta}} \\
  && \quad \quad = \frac{\Gamma(i\gamma^2)}{|\kappa|^{i\gamma^2}} \left[ \Theta(\kappa) \; e^{-\pi \gamma^2/2} + \Theta(-\kappa) \; e^{\pi   
  \gamma^2/2}\right]\;, \nonumber  \\
  &&\int_0^{\infty} d\kappa'  \;  {[\kappa']^{i\gamma^2}} \; e^{i \kappa\kappa'} = \lim_{\delta \rightarrow 1^-} 
  \frac{\Gamma(i\gamma^2+\delta)}{(-i \kappa)^{i\gamma^2+\delta}} \\
  && \quad \quad =i \; \frac{\Gamma(i\gamma^2+1)}{|\kappa|^{i\gamma^2+1}} \left[ \Theta(\kappa) \; e^{-\pi \gamma^2/2} - \Theta(-\kappa) \; e^{\pi   
  \gamma^2/2}\right]\;, \nonumber 
  \end{eqnarray}  
which hold for all $\kappa\neq 0$ from the identity 3.381(7)  of~Ref.~\onlinecite{formu} via analytic continuation [here $\Gamma(\cdots)$ is the Euler function].

Accordingly  one can easily verify that the following identity holds
  \begin{eqnarray}  
  \tilde{f}_{\rm T}^{(0)}(\kappa)  &=&  \left[ c \; \Theta(\kappa) + d \; \Theta(-\kappa)  \right]\; |\kappa|^{i \gamma^2} \;,\label{sol22} \\
  {f}_{\rm B}^{(0)}(\kappa)  &=&   -\frac{\gamma}{\kappa} \left[ c \; \Theta(\kappa) + d  \; \Theta(-\kappa)  \right]\; |\kappa|^{i \gamma^2} \;,   
  \label{sol21} 
  \end{eqnarray}
with $c$ and $d$ begin complex parameters which depend from $a$ and $b$ via the linear constraints~(\ref{constraint}). To check the  consistency  of the procedure observe  that following the same simple manipulations that led us to Eq.~(\ref{newsyst1}), from  Eq.~(\ref{fff1}) we get
  \begin{eqnarray}
  \kappa \;  \frac{d}{d \kappa} \; \tilde{f}_{\rm T}^{(0)}(\kappa) =
  i \; \gamma^2 \;\tilde{f}_{\rm T}^{(0)}(\kappa)\;, 
  \label{newsyst2}
  \end{eqnarray} 
which indeed admits the functional form of Eq.~(\ref{sol22}) as most general solution.


\begin{thebibliography}{90}
\bibitem{NOTA} 
The fields $\hat{\Psi}_\sigma({\bm r})$ in Eq.~(\ref{Eq:H0}) fulfill canonical anti-commutation relations:
%
\begin{eqnarray}
\{ \hat{\Psi}_\sigma ({\bm r}),   \hat{\Psi}^\dag_{\sigma'} ({\bm r}') \} &=& \delta_{\sigma,\sigma'} \; \delta^{(2)} ({\bm r}- {\bm r}')~,
\nonumber \\
\{ \hat{\Psi}_\sigma ({\bm r}),   \hat{\Psi}_{\sigma'} ({\bm r}')\}&=&0~.
\end{eqnarray}
%
Imposing orthonormalization conditions,
 %
 \begin{eqnarray}\label{ffd}
\int d^{2}{\bm r}\;\; \phi^*_{\sigma,n,k}({\bm r}) \; \phi_{\sigma,m,q}({\bm r}) =\delta_{n,m}\; \delta(k-q) \;,
  \end{eqnarray}
  %
  to the eigenfunctions of the unperturbed single-partilce Hamiltonian~(\ref{ham1particle})
  it then follows that 
 the fields $\hat{c}_{\sigma,n,k}$ of Eq.~(\ref{expansion}) obey the anti-commutation rules
\begin{eqnarray}
\{\hat{c}_{\sigma,n,k},  \hat{c}_{\sigma',n',k'}^\dag \} &=& \delta_{\sigma,\sigma'} \; \delta_{n,n'} \; \delta(k-k')~,
\nonumber \\
\{ \hat{c}_{\sigma,n,k},  \hat{c}_{\sigma',n',k'} \} &=& 0 \;.
\end{eqnarray}
Notice that Eq.~(\ref{ffd}) can be enforced by simply requiring that 
 \begin{eqnarray}\label{ffd1}
\int d \zeta \;\; \chi^*_{\sigma,n,k}(\zeta) \; \chi_{\sigma,m,k}(\zeta) =\delta_{n,m}\;.
  \end{eqnarray}
\end{thebibliography}
\end{document}